\input{aipcheck}

\documentclass[
    ,final            
  ]
  {aipproc}

\layoutstyle{6x9}

\usepackage{graphicx}

\begin{document}

\title{Nuclear force in Lattice QCD}


\author{Toru T. Takahashi}{
  address={Yukawa Institute for Theoretical Physics, Kyoto Univ., 606-8502, Japan}
}

\author{Takumi Doi}{
  address={RIKEN-BNL Research Center, BNL, Upton, New York 11973, USA} \\
}

\author{Hideo Suganuma}{
  address={Department of Physics, Kyoto University, Kyoto 606-8502, Japan}
}

\begin{abstract}
We perform the quenched lattice QCD analysis 
on the nuclear force (baryon-baryon interactions).
We employ $20^3\times 24$ lattice at $\beta=5.7$
($a\simeq 0.19$ fm)
with the standard gauge action and the Wilson quark action
with the hopping parameters $\kappa=0.1600, 0.1625, 0.1650$,
and generate about 200 gauge configurations.
We measure the temporal correlators
of the two-baryon system which consists of
heavy-light-light quarks.
We extract the inter-baryon force as
a function of the relative distance $r$.
We also evaluate the contribution to the nuclear force from each
``Feynman diagram'' such as the quark-exchange diagram individually,
and single out the roles of Pauli-blocking effects or quark exchanges
in the inter-baryon interactions.
\end{abstract}

\maketitle


\section{Introduction}
The nuclear force, i.e., the interaction between two nucleons,
is one of the most fundamental quantities in nuclear physics.
While its long-distance part is understood by the light-meson exchange process
proposed by Yukawa about 70 years ago,
its short-distance part has been treated phenomenologically from the
experimental data and is still mysterious at the level of quarks and gluons,
more fundamental degrees of freedom in the strong interaction.
It is naturally expected that quarks and gluons play an essential role
in the short-range interaction, which has a quite strong repulsive core
of about a few hundred MeV to 1 GeV.

In Ref.~\cite{OY},
the authors investigate the short-range nuclear force 
in the quark model using the resonating group method
and conclude that the short-range repulsive core
arises mainly from the spin-spin interactions and 
Pauli-blocking effects among quarks.

Among theoretical approaches,
the lattice QCD calculation 
is considered as the most reliable nonperturbative method.
In contrast to many studies on the hadron spectroscopy,
there have been a few lattice QCD studies on hadronic interactions.
In this paper, we show the quenched lattice QCD study of the nuclear
force, especially its repulsive core,
as a successive work of the detailed lattice-QCD studies 
of the three-quark potential~\cite{Pots1}, 
the tetra-quark potential and the penta-quark potential~\cite{Pots2}.
(The meson-meson and meson-baryon interactions
are also studied in Ref.~\cite{Doi}.)

\section{Setup and Lattice QCD Results}

One of the solid approaches for the nuclear force is 
to investigate the phase shifts
in baryon-baryon scattering processes using the L{\" u}scher formula,
but this is extremely difficult.
Instead, we study the static inter-baryon potential in lattice QCD.
To fix the center of mass of each baryon,
we use ``heavy-light-light'' quark system.
Here, the heavy quark is treated as a static one with infinite mass.
Then, the inter-baryon distance can be clearly
defined as the relative distance between two heavy (static) quarks.
For the interpolating field of this heavy-light-light quark baryon,
we employ
$
N(\vec r, t)\equiv \varepsilon_{abc}
Q^a(\vec r, t)\left[^tq_1^b(\vec r, t)C\gamma_5q_2^c(\vec r, t)\right],
$
with $Q^a(\vec r, t)$ the field for a static quark located
at $(\vec r, t)$ and $q_i^a(\vec r, t)$ the light-quark field.
The inter-baryon potential $V_{BB}(r)$ 
as a function of the relative distance $r\equiv |\vec r|$
can be extracted from the temporal correlators,
$
C_{BB}(\vec r, T)\equiv
\langle [N(\vec 0, T)N(\vec r, T)]
[\bar N(\vec 0, 0)\bar N(\vec r, 0)]\rangle,
$
and its limit as $\lim_{T\rightarrow \infty}-\frac{1}{T}\ln C_{BB}(\vec r, T)$.

The correlators can be expressed as
the sum of the products of six quark propagators
via the Wick contraction of the quark fields.
In particular, the propagator for the static quarks
is expressed as the path-ordered product of the gauge field
${\rm P}\exp(ig\int A_0(x)dt)$,
which corresponds to the leading-order propagator 
in the heavy quark approximation.
\begin{figure}[h]
\includegraphics[scale=0.4,trim=0 460 0 -60, clip]{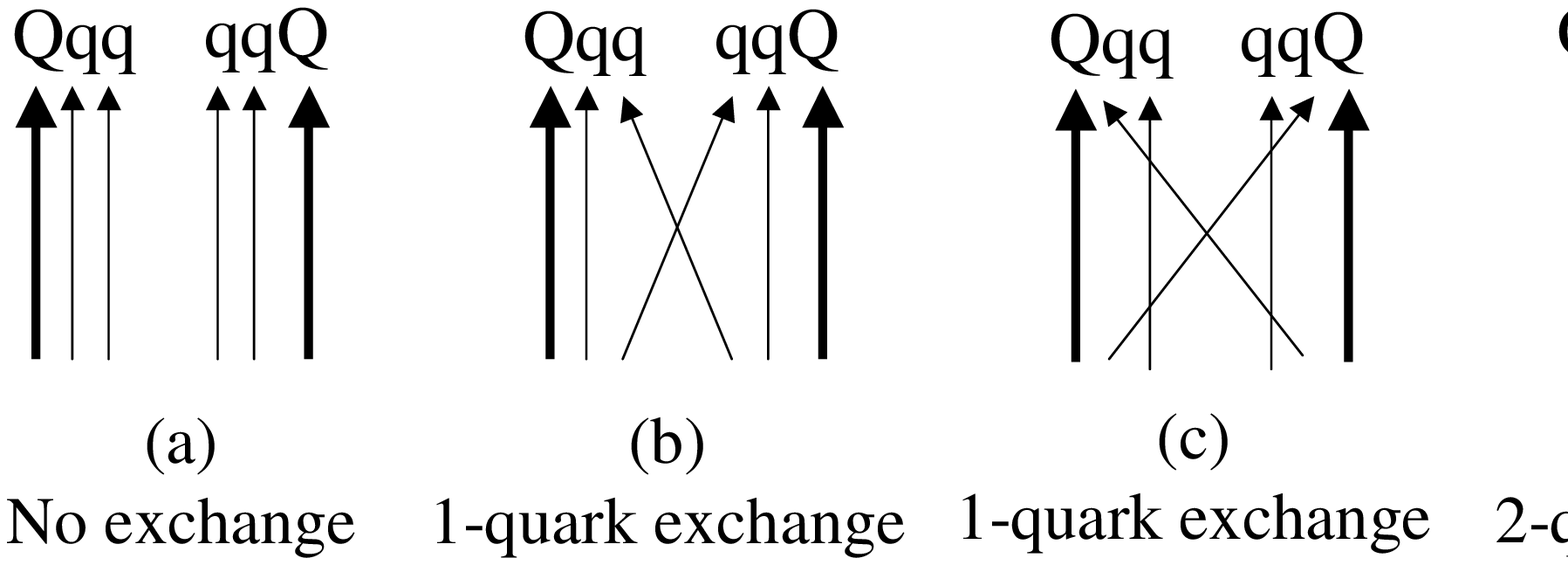}
\caption{
\label{diagrams}
Schematic figures of the Wick contraction.
}
\end{figure}
As for the light quarks,
the flavor content can be controlled by selecting ``Feynman diagrams''.
For example, in the case when all the quarks have different flavors,
we omit the quark-exchange diagrams.
(We need only (a) in Fig.~\ref{diagrams}.)
If some pairs of quarks are identical, 
we include the corresponding exchange diagram of the two quarks
((b)-(d) in Fig.~\ref{diagrams}).
In such a way, we can control the flavor content,
which is directly connected to the Pauli-blocking effects among quarks.

Note here that
one of quarks is static and even the ``light'' quarks are 
rather heavy as $m_{\rm u,d}^{\rm current}\simeq 100\sim 250$ MeV,
which would weaken the Pauli-blocking effects and
the spin-spin interactions proportional to $1/m_{\rm const}^2$,
with $m_{\rm const}$ the constituent quark mass.
However, taking into account that the short-range interactions between
two nucleons are quite strong,
the present setup would be enough to single out the essence
of the repulsive core in the nuclear force.

\vspace{-.1cm}
\begin{figure}[h]
\includegraphics[scale=0.3]{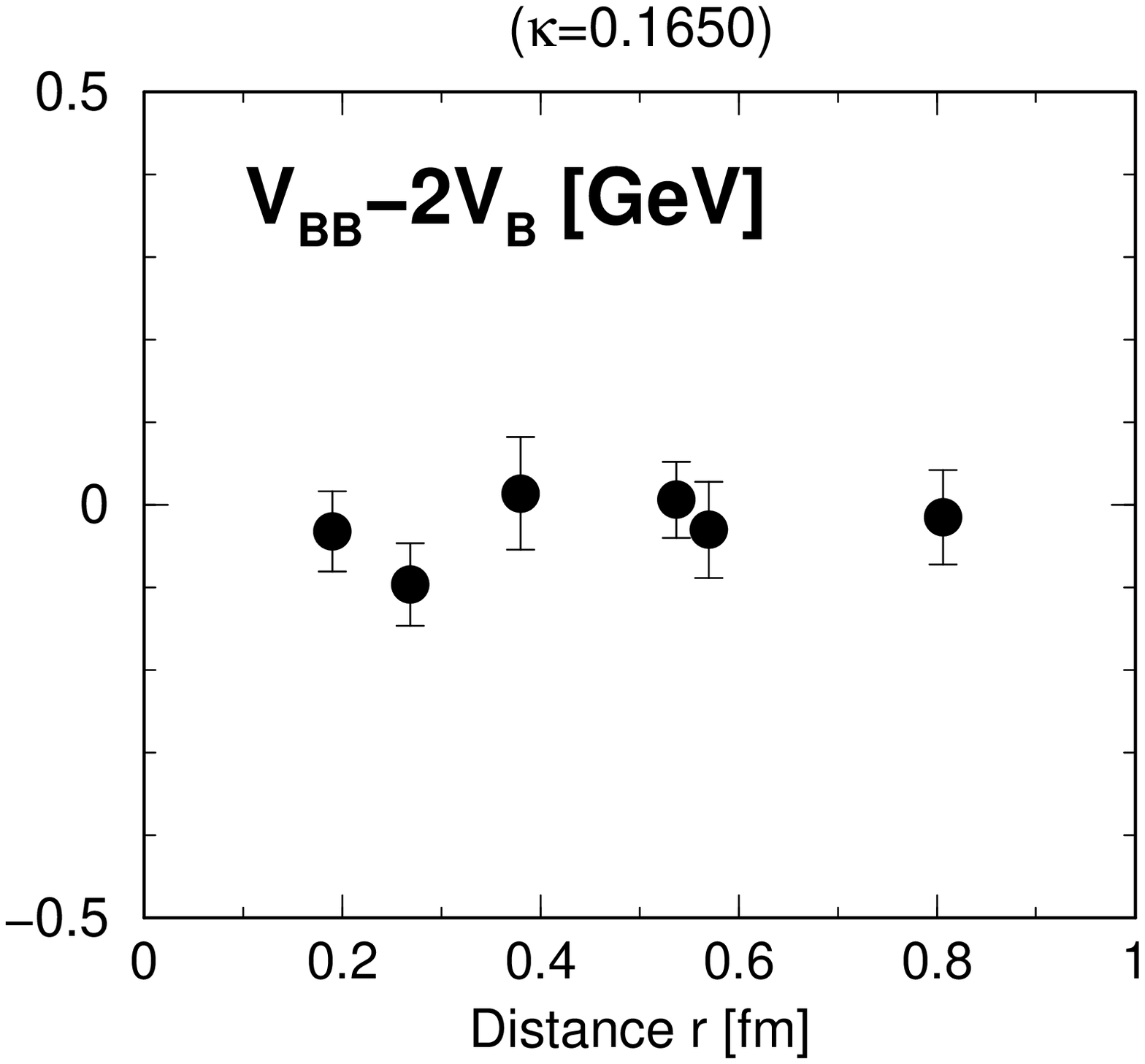}
\includegraphics[scale=0.3]{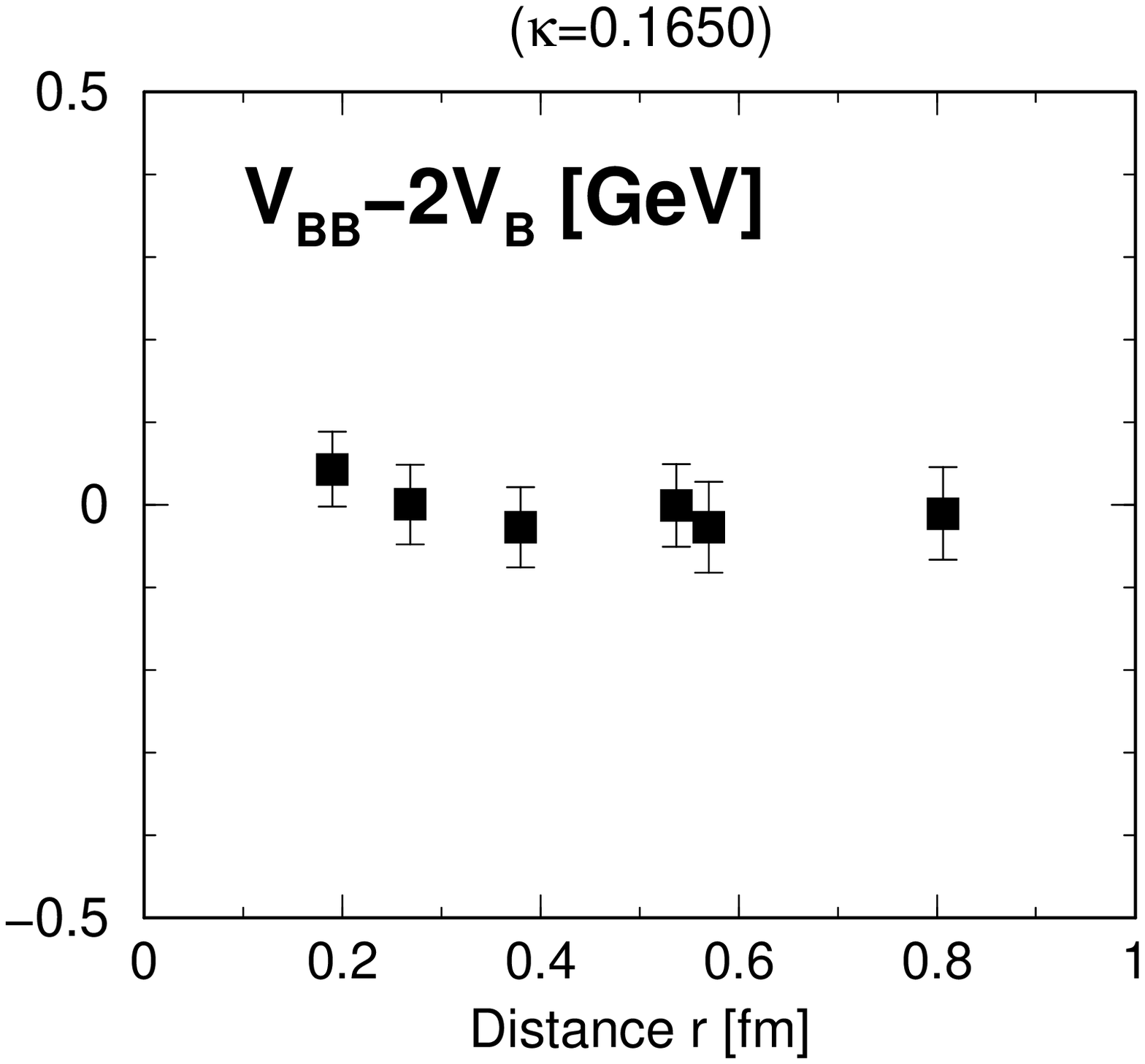}
\includegraphics[scale=0.3]{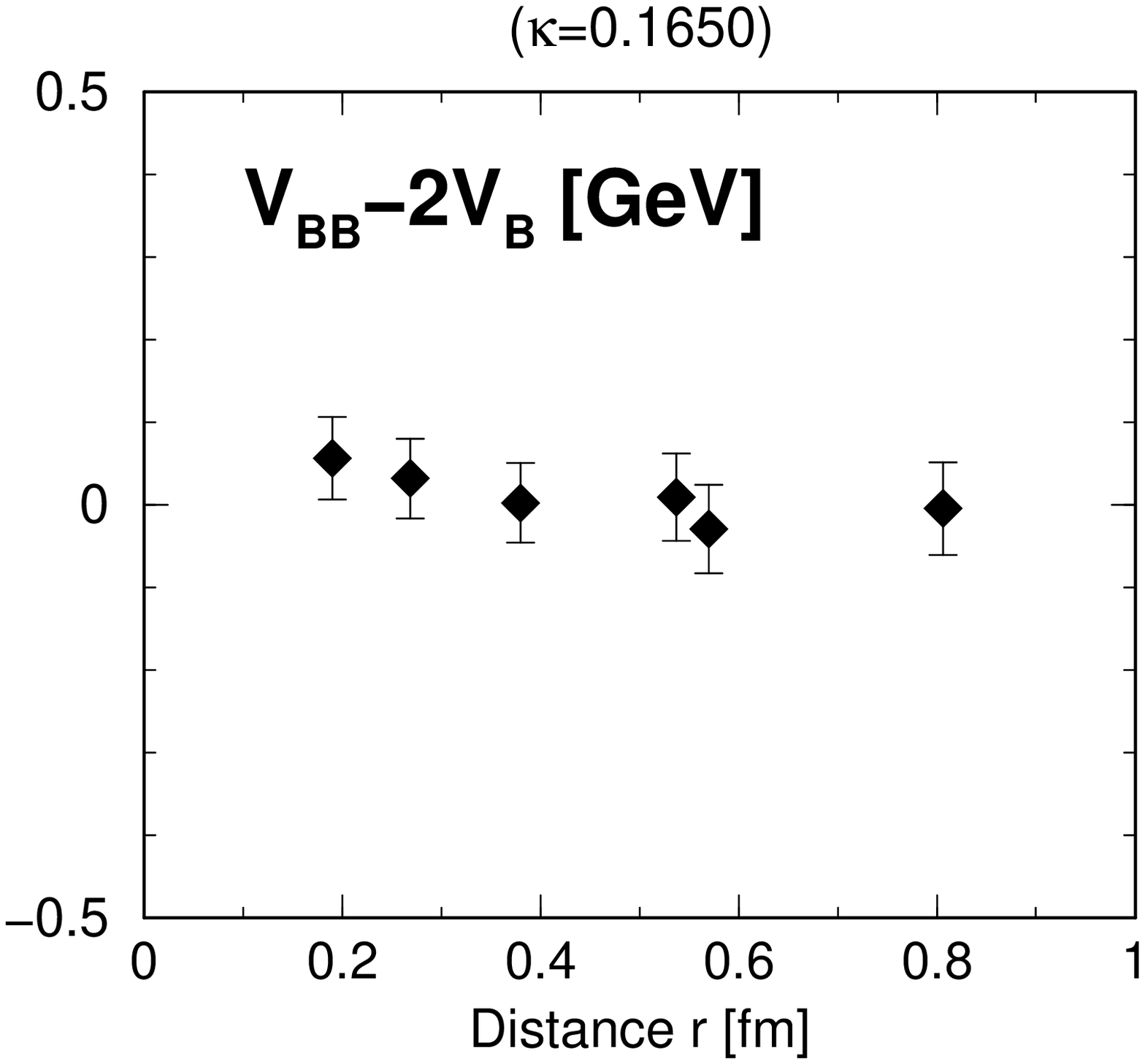}
\caption{
The inter-baryon potential $V_{BB}(r)-2V_B$ in lattice QCD.
{\it Left}:All the light-quark flavors are different.
{\it Middle}:One pair of quark flavors are identical.
{\it Right}:Two pairs of quark flavors are identical.
The horizontal axis denotes the inter-baryon distance $r$.
\label{results}
}
\end{figure}
We show the lattice QCD data of $V_{BB}(r)-2V_B$
as a function of the relative distance
$r$ in Fig.~\ref{results},
where $V_B$ is the energy of the single heavy-light-light baryon.
The left figure shows the result when all the light-quark flavors are different.
The middle and right figures indicate the results
when one or two pairs of quark flavors are identical, respectively.

Surprisingly, we observe almost no repulsive core even in the case
when the distance between two nucleons is about 0.2 fm,
and find that the interaction seems very weak in the whole range
of $r$.
It seems puzzling since the expected ``short-range repulsive force''
of hundreds MeV could be detected in our setup.

One possible reason is that two nucleons deform such that
they avoid the strong repulsive force.
An extreme possibility is that the ground state we have observed
may be not a heavy-light-light plus heavy-light-light (Qqq+Qqq) system
but a heavy-heavy-light plus light-light-light (QQq+qqq) system.
If this is the case, the ground-state energy of the system
cannot exceeds $m({\rm QQq})+m({\rm qqq)}$.
In other words, 
when the energy of Qqq+Qqq system is larger than
the ``threshold'' $m({\rm QQq})+m({\rm qqq})$,
light baryon (qqq) decouples such that the system energy is lowered.

To prevent the recombination processes, it may be useful
to enclose the two-baryon system 
in a sub-lattice $\Gamma$ of $(L_x,L_y,L_z)$
with the Dirichlet boundary, $q(x)|_{\partial \Gamma}=0$.
In this setup, the decoupled light baryon inevitably
has a finite momentum such as $(\pi/L_x,\pi/L_y,\pi/L_z)$,
and the ``threshold''
$m({\rm QQq})+m({\rm qqq})$ is raised up,
which suppresses the recombination to QQq+qqq.
(We have computed the inter-baryon potential with the small Dirichlet box.
However, we still find no strong repulsive force.)
We note that the behavior of the spectral weights would
also clarify whether it is the case or not.

\section{Summary}
We have studied the nuclear force
using quenched lattice QCD.
We have extracted the inter-baryon potential as
a function of the relative distance $r$
and have also evaluated the contribution to the nuclear force from each
Feynman diagram such as the quark-exchange diagram individually.
As a result, we have observed no significant repulsive core
in the baryon-baryon interactions,
in spite of whether the quark-exchange processes
(Pauli-blocking effects among quarks) are included or not.
In any case, the results are rather nontrivial and
may cast light on the internal hadron structure 
and the inter-hadron interactions.
We are trying several methods 
in order to get more solid results
and to discover the nature of the nuclear force at the quark-gluon level.


The lattice QCD Monte Carlo calculations have been performed
on NEC-SX5 at Osaka University and on HITACHI-SR8000 at KEK.

\end{document}